\documentclass{raa}

\usepackage[varg]{txfonts}

\usepackage{natbib}
\usepackage{graphicx}
\usepackage{epstopdf}
\usepackage{amssymb}
\usepackage{float}

\bibpunct{(}{)}{;}{a}{}{,}

\begin{document}

\title{Meta-analysis of Electron Cyclotron Resonance Absorption Features Detected in High-Mass X-ray Binaries}

   \volnopage{Vol.0 (200x) No.0, 000--000}      %%preserved for Editor. DOn't remove!
   \setcounter{page}{1}          %%starting page, preserved for Editor. DOn't remove!

\author{
Dimitris M. Christodoulou\inst{1,2},
Silas G. T. Laycock\inst{1,3},
\and
Demosthenes Kazanas\inst{4},
}
%%%

\institute{
Lowell Center for Space Science and Technology, University of Massachusetts Lowell, Lowell, MA, 01854, USA\\
\and
Department of Mathematical Sciences, University of Massachusetts Lowell, Lowell, MA, 01854, USA.
Email: dimitris\_christodoulou@uml.edu\\
\and
Department of Physics \& Applied Physics, University of Massachusetts Lowell, Lowell, MA, 01854, USA.
Email: silas\_laycock@uml.edu, rigelcappallo@gmail.com \\
\and
NASA Goddard Space Flight Center, Laboratory for High-Energy Astrophysics, Code 663, Greenbelt, MD 20771, USA. Email: demos.kazanas@nasa.gov \\
}

\date{Received~~2019 month day; accepted~~2019~~month day}

\abstract{
Using recent compilations of detailed X-ray observations and spectral models of exceptional quality, we record the electron cyclotron resonance absorption (ECRA) features that have been detected in 45 pulsating high-mass X-ray binaries (HMXBs) and ultraluminous X-ray (ULX) sources harboring neutron stars, although seven of these detections are still questionable and another 21 are single and/or not independently confirmed. From the comprehensive catalogs of Jaisawal \& Naik and Staubert et al. and from several additional recent observations, we produce two lists of HMXB ECRA sources: a list of 17 sources in which multiple ECRA lines or single very low-energy lines are seen, in which we can reasonably assume that the lowest energy reveals the fundamental cyclotron level for each source; and a ``contaminated'' list of 38 sources including the 21 detections of single ECRA lines that may (not) be higher-level harmonics. Both lists confirm a previous result that we have obtained independently by modeling the propeller lines of Magellanic HMXB pulsars: the surface dipolar magnetic fields $B_*$ of HMXB neutron stars are segregated around five distinct values with $B_* = 0.28\pm 0.08, 0.55\pm 0.11, 1.3\pm 0.37, 3.0\pm 0.68$, and $7.9\pm 3.1$, in units of TG. An explanation of this phenomenon is currently lacking. We have found no correlation between these $B_*$ values and the corresponding observed spin periods, spin period derivatives, orbital periods, maximum X-ray luminosities, neutron star masses, or companion star masses.
}

\authorrunning{Christodoulou et al.}
\titlerunning{Electron Cyclotron Resonance Absorption Features in HMXBs}

\maketitle

\section{Introduction}\label{intro} 

Based on the aggregate works of \cite{jai17} and \cite{sta19}, we have compiled tables of electron cyclotron resonance absorption (ECRA) features and the inferred surface dipolar magnetic fields in (extra)galactic high-mass X-ray binaries (HMXBs) and ultraluminous X-ray (ULX) sources. Our work was motivated by a recent claim that a \textit{proton} cyclotron absorption feature was detected in the X-ray spectrum of a ULX source in M51 \citep{bri18}, a claim that has since been clearly refuted for dipolar fields \citep{mid19}. We did not believe the original expos\'e in the first place because cyclotron power scales as $1/m^2$ \citep[][Section 14.2, pages 469-471]{jac62}, where $m$ is the mass of the spiralling particle around magnetic field lines. Since the proton-to-electron mass ratio is 1836, the emissivity of gyrating protons is smaller by a factor of $3.37\times 10^6$, thus electrons are widely believed to dominate the power spectra of both Galactic and extragalactic X-ray sources. 

Cyclotron resonance features are used to determine the strength of the surface magnetic field $B_{\rm cyc}$ of the compact object. Were they due to protons, then $B_{\rm cyc}$ would appear to be stronger by a factor of $\approx 1836$ (for the same X-ray luminosity), making the field of magnetar strength. This is not what is observed in Magellanic HMXBs \citep{chr16,chr17} and in the ULX N300 X-1 \citep{wal18,chr18a}, so we do not subscribe to the notion of dipolar $B_*$ stellar values above the quantum limit of $44.14$~TG. Furthermore, there is independent evidence that HMXBs and ULXs do not exhibit magnetar-strength magnetic fields \citep{kin19}.

For a fundamental electron cyclotron line centered at energy $E_{\rm cyc}$ and for the canonical pulsar gravitational redshift of $z_g=0.306$, the dipolar surface magnetic field $B_{\rm cyc}$ is determined from the equation \citep{chr18c}
\begin{equation}
B_{\rm cyc} = \left(\frac{E_{\rm cyc}}{8.86~{\rm keV}}\right)  ~{\rm TG} \, .
\label{cyc}
\end{equation}
Table~\ref{table1} shows our main working list of known ECRA lines $E_{\rm cyc}$ and their implied $B_{\rm cyc}$ values taken from the comprehensive compilations of \cite{jai17} and \cite{sta19}. These values are the lowest observed lines in sources with multiple energy levels, except for Group C in which the lines are single but they are too low in energy to doubt that they may not represent the corresponding fundamental energy levels \citep{mid19,mai18,wal18}.

We note that, as in the case of SXP15.3 (Table~\ref{table1}), M51 ULX8 falls exactly on to the so-called second Magellanic propeller line \citep{chr17}. Under the circumstances, we proceed with the hypothesis that dipolar magnetic fields in neutron-star HMXBs as well as neutron-star ULX sources do not take extreme values. In \S~\ref{results}, we present our detailed clustering analysis of pulsar magnetic fields from the known ECRA lines and, in \S~\ref{conc}, we summarize and discuss our results.

\begin{table}
\caption{Fundamental Cyclotron Absorption Lines and Surface Dipolar Magnetic Fields of NSs in HMXBs}
\label{table1}
\centering
\begin{tabular}{llll}
\hline
No. & X-ray Source & $E_{\rm cyc}$ & $B_{\rm cyc}$  \\
      &            & (keV)              & (TG)               \\
\hline
\multicolumn{4}{c}{Group A: Confirmed Lines$^{(a)}$} \\
\hline
1    &  4U 0115+63      &  12                 &    1.35            \\
2    &   4U 1907+09   &   18                  &    2.03             \\
3    &  4U 1538-52     &  22                   &   2.48                     \\
4    &   Vela X-1          &  25                   &   2.82                     \\
5    &   V 0332+53     &  27                   &   3.05                     \\
6    &   Cep X-4          &  28                   &   3.16                     \\
7    &   MAXI J1409-619 &  44               &   4.97                     \\
8    &   1A 0535+262     &  45               &   5.08                     \\
9    &  1A 1118-616    &  55                   &    6.21            \\
\hline
\multicolumn{4}{c}{Group B: Tentative Lines$^{(a)}$} \\
\hline
10  &  Swift 1626.6-5156             & 10    &   1.13   \\
11  &  XMMU 054134.7-682550   &  10    &   1.13   \\
12  &  EXO 2030+375                 &  11    &   1.24   \\
13  &  IGR 17544-2619               &  17    &   1.92    \\
14  &  2S 0114+650                   &  22     &   2.48    \\
\hline
\multicolumn{4}{c}{Group C: Low-Energy Extragalactic Lines$^{(b)}$} \\
\hline
15  &  M51 ULX8       &  4.5   & 0.508     \\
16  &  SXP15.3          &   5.0  & 0.564      \\
17  &  NGC300 ULX1  &  13   &  1.47      \\

\hline
\end{tabular}
\\
($a$)~From Table 1 of \cite{jai17} and Table 4 of \cite{sta19}. ($b$)~From \cite{mid19}, \cite{mai18}, and \cite{wal18}, respectively.
\end{table} 

\section{Results}\label{results}

\subsection{Table~\ref{table1} Data}

The first 14 entries in Table~\ref{table1} are HMXB sources for which we ran a cross-comparison between the lists of \cite{jai17} and \cite{sta19}. In each case, multiple ECRA lines have been observed and most of them have been confirmed by follow-up observations. The listed values of $E_{\rm cyc}$ are the lowest in each case and we believe that they represent the fundamental energies in these sources from which we can infer the surface magnetic fields $B_{\rm cyc}$ of the neutron stars. The additional entries 15-17 (Group C) represent single detections of extragalatic X-ray sources \citep[][respectively]{mid19,mai18,wal18} that are so low to make us assume that they also represent the fundamental ECRA energies in these objects.

\begin{figure}
\includegraphics[scale=0.6]{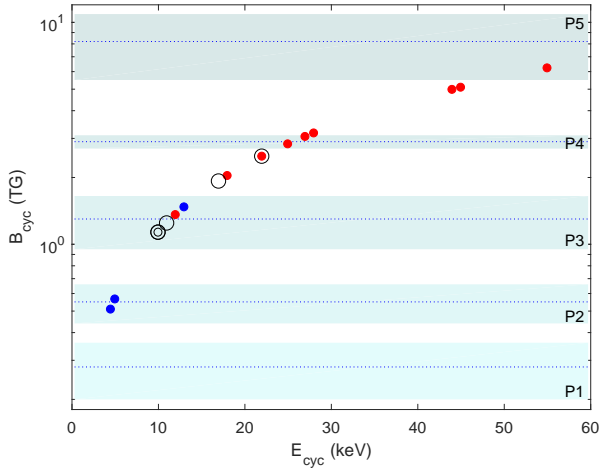}
\caption{Observed fundamental energy levels $E_{\rm cyc}$ and the inferred magnetic fields $B_{\rm cyc}$ for the X-ray sources listed in Table~\ref{table1}. The five propeller lines found in Magellanic HMXBs are shown by dotted lines and their spreads in $B_{*}$ stellar values are shaded and numbered as P1 to P5 from the lowest to the highest line. Key (based on Table~\ref{table1}): Red dots $\to$ Group A; Open circles $\to$ Group B; Blue dots $\to$ Group C. Different circle sizes are used to accommodate coincident points. The only outliers appear to be the points at $E_{\rm cyc}=17$~keV and 18~keV that fall in the middle of the gap between areas P3 and P4.
}
\label{fig1}
\end{figure}

\begin{figure}
\includegraphics[scale=0.6]{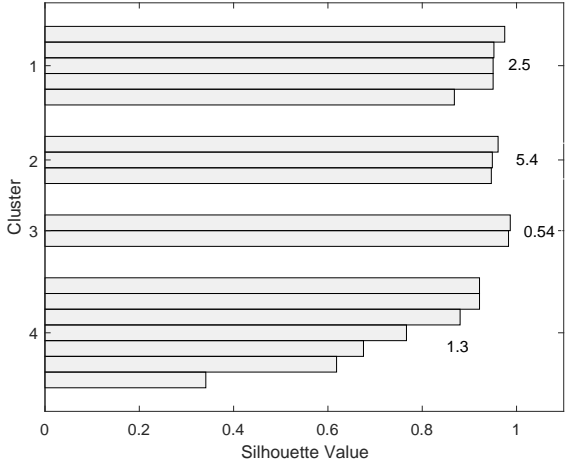}
\caption{Silhouette diagram of the data shown in Figure~\ref{fig1}. The centroids of the clusters are also shown in units of TG. Clustering analysis shows only one potential outlier with a silhouette value of SV$< 0.6$, the bottom point in cluster 4 with $E_{\rm cyc}=18$~keV.
}
\label{fig2}
\end{figure}

\begin{figure}
\includegraphics[scale=0.6]{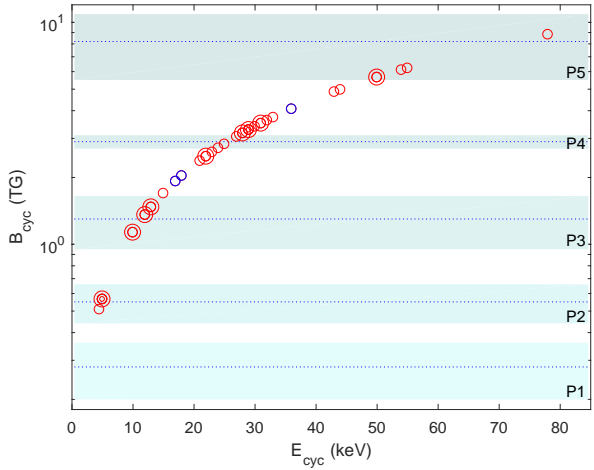}
\caption{Observed lowest-energy levels $E_{\rm cyc}$ and the inferred magnetic fields $B_{\rm cyc}$ for the HMXB sources listed in Tables 4 and 5 of \cite{sta19}. All unquestionable detections have been included, and single-level detections are assumed to measure the fundamental cyclotron energies of such sources. The five propeller lines and their shaded areas are as in Figure~\ref{fig1}. Different circle sizes are used to accommodate coincident points. The three outliers are marked by blue circles ($E_{\rm cyc}=17$~keV, 18~keV, and 36~keV).
}
\label{fig3}
\end{figure}

\begin{figure}
\includegraphics[scale=0.6]{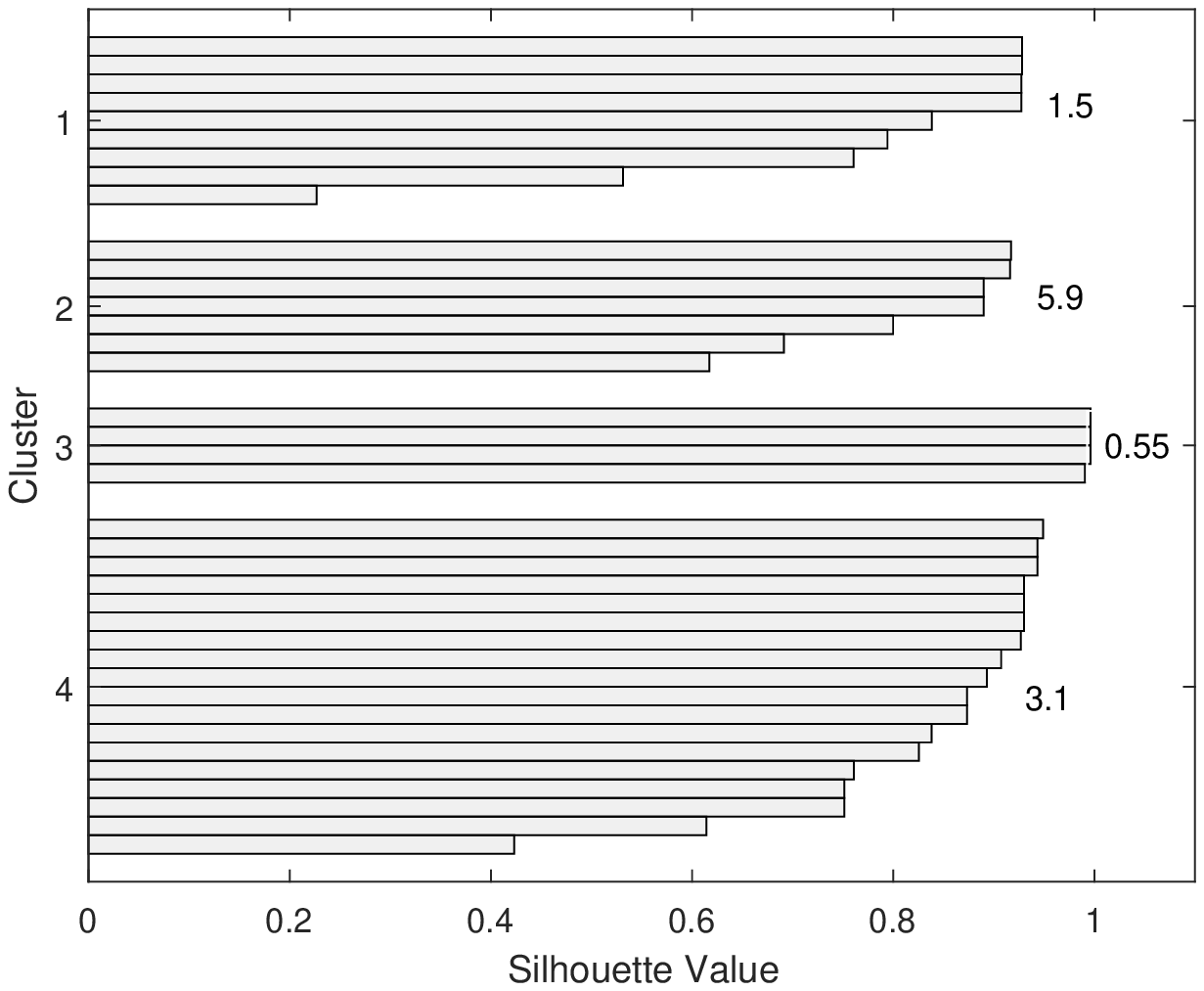}
\caption{Silhouette diagram of the data shown in Figure~\ref{fig3}. The centroids of the clusters are also shown in units of TG. Clustering analysis shows only three potential outliers with SV$<0.6$, two in cluster 1 and one in cluster 4 (blue circles in Figure~\ref{fig3}).
}
\label{fig4}
\end{figure}

Figure~\ref{fig1} shows the $E_{\rm cyc}$-$B_{\rm cyc}$ plane for the sources listed in Table~\ref{table1}. The five propeller lines found in the Magellanic HMXBs \citep{chr16,chr17} are shown by dotted lines and their spreads in $B_{\rm cyc}$ values are shaded and numbered as P1 to P5 from the lowest to the highest and most uncertain propeller line. The magnetic fields determined from ECRA lines (eq.~(\ref{cyc})) segregate within the propeller areas P2 to P5. This is not a surprising result since neutron stars and their surface magnetic fields share quite a few common physical properties. By visual inspection, Figure~\ref{fig1} appears to exhibit only two outliers (in the gap between the shaded areas P3 and P4) at  $E_{\rm cyc}=17$~keV and 18~keV. A rigorous clustering analysis \citep{rou87,kau90} indicates that only one of these points ($E_{\rm cyc}=18$~keV) is a true outlier with a silhouette value of SV$ < 0.6$ (Figure~\ref{fig2}). In this analysis, we used the K-means algorithm \citep{seb84,spa85} with squared Euclidean distances to assign cluster membership in the clusters shown in Figure~\ref{fig2}. A value of SV$> 0.6$ is usually considered sufficient for definitive cluster membership with no overlap between neighboring clusters. Furthermore, clusters 1-3 in Figure~\ref{fig2} show an extremely dense composition with SV values $> 0.9$. This result constitutes an independent confirmation of the multiple propeller lines found in a study of Magellanic HMXB pulsars \citep{chr17}.

\subsection{All Unquestionable HMXB data in \cite{sta19}}

We have also considered the entire list of the 38 unquestionable HMXB ECRA lines compiled by \cite{sta19} (their Tables 4 and 5), including single and/or unconfirmed detections. The corresponding distribution of magnetic fields is shown in Figure~\ref{fig3} and the results from a formal clustering analysis are illustrated in Figure ~\ref{fig4}. This figure indicates that there are only three potential outliers with SV$<0.6$, two in cluster~1 and one in cluster~4 (also indicated by blue circles in Figure~\ref{fig3}). We note that the three outliers cannot change their cluster membership because their SVs are positive. The remaining 35 sources are all densely packed (with SV$> 0.6$) within their respective clusters, as shown in Figure~\ref{fig4}.

\begin{table*}
\caption{Segregated $B_{*}$ (TG) Values and Ranges in HMXB Neutron Stars}
\label{table2}
\centering
\begin{tabular}{l|ll|ll|ll}
\hline
Propeller & Magellanic&K-means & Figure~\ref{fig3} &K-means & Combined &Combined \\
Line & Sources$^{(a)}$ &Centroids& Data Set$^{(b)}$ &Centroids&Data Sets&Ranges\\
\hline
P1 & $0.28\pm 0.08$   &0.286& \dots                  &\dots&  $0.28\pm 0.08$  & [0.20, 0.36] \\
P2 & $0.55\pm 0.11$   &0.528&  $0.54\pm 0.03$ &0.550&  $0.55\pm 0.11$ & [0.44, 0.66] \\
P3 & $1.3\pm 0.35$     &1.20&  $1.4\pm 0.28$    &1.48&  $1.3\pm 0.37$     & [0.95, 1.7] \\
P4 & $2.9\pm 0.20$     &2.97&  $3.0\pm 0.68$    &3.10&  $3.0\pm 0.68$     & [2.3, 3.7] \\
P5 & $8.2\pm 2.7$       &8.04&  $6.8\pm 2.0$      &5.92&  $7.9\pm 3.1$       & [4.8, 11] \\

\hline
\end{tabular}
\\
($a$)~From Figure 2 of \cite{chr17}. $(b)$~From Tables 4 and 5 of \cite{sta19}.
\end{table*}

In Table~\ref{table2}, we combine the $B_{\rm cyc}$ measurements from the data in \cite{sta19} and the $B_*$ measurements from \cite{chr17} and their spreads. The combined results indicate a clustering of dipolar surface magnetic-field values around the values of $B_* = 0.28\pm 0.08, 0.55\pm 0.11, 1.3\pm 0.37, 3.0\pm 0.68$, and $7.9\pm 3.1$, in units of TG. The overall range of magnetic-field values is 0.20-11~TG, corresponding to 0.45-25\% of the quantum limit of 44.14~TG.

\section{Summary and Discussion}\label{conc}

\subsection{Summary}

We have analyzed the comprehensive compilations of multiple ECRA lines presented by \cite{jai17} and \cite{sta19} for (extra)galactic HMXBs and we have also included additional cyclotron-line results from \cite{mid19}, \cite{mai18}, and \cite{wal18} for M51 ULX8, SXP15.3, and N300 ULX1, respectively. These results allow us to calculate the dipolar surface magnetic fields of these sources under the assumption that the lowest-energy lines represent the fundamental cyclotron levels (Figures~\ref{fig1} and~\ref{fig3}). Table~\ref{table1} includes 17 X-ray sources in which multiple lines have been observed or the detected single lines are very low (Group C: 4.5-13~keV), and it is reasonable to assume that they all capture the fundamental cyclotron level in each X-ray source.

We calculated the surface dipolar magnetic fields $B_{\rm cyc}$ (eq.~(\ref{cyc})) from these cyclotron lines, and we found that their values match the multiple propeller lines that we have found independently from an empirical study of Magellanic HMXB pulsars (Table~\ref{table2}). We ran a formal clustering analysis of the cyclotron data (Figures~\ref{fig2} and~\ref{fig4}) that leaves no doubt that the $B_{\rm cyc}$ values populate dense clusters centered around the values of $B_* = 0.28, 0.55, 1.3, 3.0,$ and 7.9~TG, just as in the case of Magellanic HMXB pulsars. This result is not surprising---neutron stars share many physical properties and their surface magnetic fields could not possibly be randomly distributed since these compact objects are produced by the same physical process. Modest variations in their structural properties \citep{lat01} could not possibly produce too different anchored dipolar magnetic fields.

\begin{figure}
\includegraphics[scale=0.6]{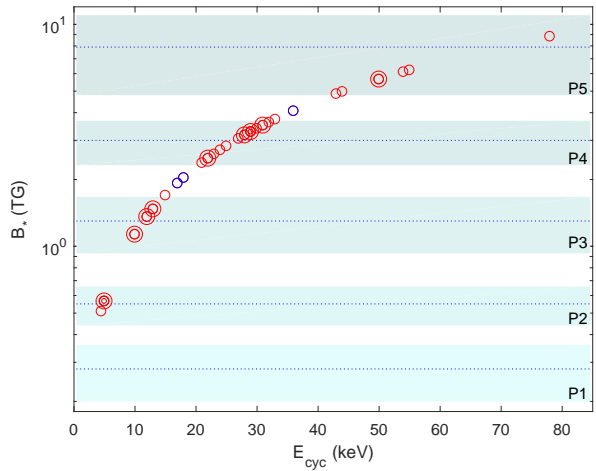}
\caption{As in Figure~\ref{fig3}, from the data in Tables 4 and 5 of \cite{sta19}, but the spreads of the five propeller lines (shaded areas) have been updated as a result of this work. Different circle sizes are used to accommodate coincident points. The three clear outliers (out of 38 points) are marked by blue circles ($E_{\rm cyc}=17$~keV, 18~keV, and 36~keV).
}
\label{fig5}
\end{figure}

Figure~\ref{fig5} shows the aggregate result including the results of the present investigation. This figure shows that there are only 3 outliers (denoted by blue circles) in the group of 38 sources (8\%), 21 of which are uncertain, but nevertheless most of them fall within the P2-P5 shaded areas determined by this study: $B_* =  0.55\pm 0.11, 1.3\pm 0.37, 3.0\pm 0.68$, and $7.9\pm 3.1$, in units of TG. We note that the spread of area P4 ($\pm 0.68$), which was extremely tight in Magellanic sources, has now more than tripled owing to the \cite{sta19} data used in this study. All other areas have expanded by very small amounts or not at all.

\subsection{Discussion}

\begin{figure}
\includegraphics[scale=0.6]{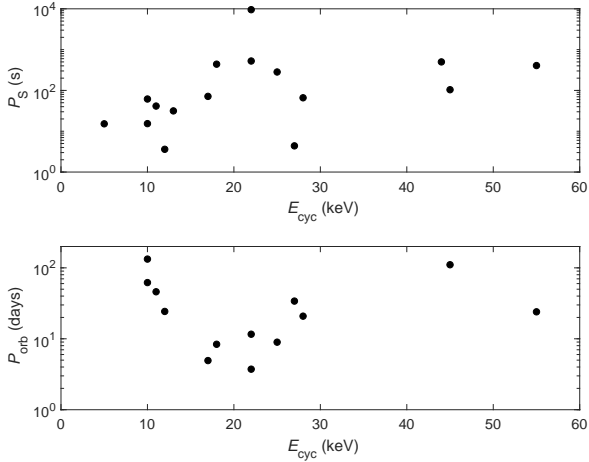}
\caption{The spin periods $P_{\rm S}$ and the orbital periods $P_{\rm orb}$ are plotted versus $E_{\rm cyc}$ for the sources listed in Table~\ref{table1}. Data are taken from the references listed in Table~\ref{table1}.
}
\label{fig6}
\end{figure}

The observed segregation of magnetic fields in HMXB and ULX pulsars requires a physical origin that is currently lacking. We searched for correlations between the observed ECRA energies $E_{\rm cyc}$ and various other measured physical properties in these systems. Most of the data that we used are listed in Tables 4, 5, 6, and 8 of \cite{sta19}. We were unable to find any trend or correlation between $E_{\rm cyc}$ (equivalently $B_{\rm cyc}$) and the observed spin periods, spin period derivatives, binary orbital periods, maximum X-ray luminosities, neutron star masses, or companion star masses. We show two representative examples in Figure~\ref{fig6}, where we plot the spin periods $P_{\rm S}$ and the orbital periods $P_{\rm orb}$ versus the observed cyclotron-energy values $E_{\rm cyc}$. No discernible trend is seen in these plots. This leaves us presently in the dark and we can only speculate as follows.

The magnetic field distribution (MFD) of neutron stars has long been assumed to 
descend directly from the MFD of their precursors, OB and WR stars. However, only recently have statistically meaningful MFDs for massive stars been assembled, principally by
the MiMeS \citep{wad16} and BOBs \citep{fos16} collaborations. A synthesis and meta-analysis of these results by \cite{med17} has reported the following salient points: The intrinsic fraction of magnetic stars is 6-7\%, the rate of field decay is a strong function of mass, and the magnetic fields $B$ of OB stars obey a log-normal distribution centered on a mean of $\langle\log [B/(1\,{\rm kG})]\rangle = -0.5$ with a standard deviation of 0.5. No evidence was found for any discrete features in this distribution. Comparison with the MFD of normal radio pulsars (their figure 7) revealed that the total magnetic flux ${\it\Phi}$ in neutron stars is about 3 orders of magnitude lower than that in their progenitors, which indicates a surprising non-conservation of ${\it\Phi}$ in HMXBs. The MFD for WR stars lies between these two populations, in which case field decay during the final stages of stellar evolution could be the source of the discrepancy, possibly rescuing the principle of flux conservation during the supernova explosions of such massive stars.

The clustering of magnetic field values reported here and in \cite{chr17} arises from independent data sets and two different methodologies, so we can be confident that it is real and that there must be a physical explanation for this phenomenon. The smoothness of the MFD for OB stars \citep{med17} does not obviously provide an intrinsic set of preferred magnetic field values in OB stars. But magnetic field decay during the evolutionary stage leading up to the supernova explosion could provide a mechanism to break this smoothness. We speculate that binary interaction affects the rate of magnetic field decay. It is plausible that some type of spin-orbit coupling could imprint some preferred values of the magnetic flux ${\it\Phi}$ shortly before the explosion, after which the onset of collapse would amplify the surface magnetic field $B_*$ of the resulting neutron star just as expected.

\begin{acknowledgements}
This work was supported in part by NASA ADAP grants NNX14-AF77G and 80NSSC18-K0430.
\end{acknowledgements}

\label{lastpage}

\end{document}